\documentclass[prb,twocolumn]{revtex4-2}
\usepackage{graphicx}
\usepackage[usenames,dvipsnames]{color}

 \usepackage{amsmath}
\usepackage{xcolor}
\begin{document}

\title{Creation, annihilation and transport of nonmagnetic antiskyrmions within Ginzburg-Landau-Devonshire model  }

\author{V. Stepkova and J. Hlinka}
\affiliation{FZU - Institute of Physics of the Czech Academy of Sciences\\%
Na Slovance 2, 182 21 Prague 8, Czech Republic}

\date{\today}

\begin{abstract}
Atomistic model calculations recently revealed that the polarization pattern at the cross-section through  nanoscale ferroelectric nanodomains in rhombohedral barium titanate
shows a clear antiskyrmion attributes such as the net invariant topological charge of minus two.
The present work confirms that these topological defects also exist as local minima in the  discretized Ginzburg-Landau-Devonshire model, which is widely used in the phase-field modeling studies. 
We explore by phase-field simulations how an electric field can be used to create
or move antiskyrmions within a monodomain ferroelectric matrix.
The process of the field-induced antiskyrmion annihilation reveals the existence of a discrete ladder of intermediate antiskyrmion states with distinct radii but common symmetry, topology and geometrical properties.
\end{abstract}


\maketitle 

\section{Introduction}

The interest in nanoscale topological defects such as vortices and skyrmions is an important driving factor for the progress of current ferroelectric physics and materials science\cite{Das2019, WangYJ2023,  Junquera2023, Nataf2020, ZhuR2022, Guo2022, Halcrow2023, Halcrow2024, Prokhorenko2024}.
Very recently, a peculiar type of topological defect, so called antiskyrmion, was found in the groundstate rhombohedral phase of the classical ferroelectric perovskite, barium titanate\cite{Goncalves2024}.
More precisely, computer simulations using an ab-initio based atomistic model, the shell model of Refs.\,\onlinecite{Tinte2004,Sepliarsky2005}, have shown that the antiparallel columnar nanodomains in this phase are spontaneously decorated by a particular non-collinear polarization texture, which gives rise to an unusual skyrmion number of -2 per each such nanodomain\cite{Goncalves2024}.
The relatively thick non-collinear boundary region separating the  monodomain ferroelectric background and the inverted nanodomain core could indicate an overall high mobility of this boundary and its shape instability.
Instead, it has been found that once the regular antiskyrmion is formed, it is rather stable in its size, shape and position at least up to temperatures of several tens of Kelvins\cite{Goncalves2024}.
It has been also inferred that the formation of the antiskyrmion is related to the unusual anisotropy of the local and nonlocal interactions that are typical for certain rhombohedral ferroelectric perovkites, like barium titanate and potasium niobate\cite{Goncalves2024}.
Considering the surprising geometry and topology of these antiskyrmionic polarization patterns, it is highly desirable to seek their confirmation  within a radically different theoretical framework.

In the present work we thus explored the properties of antiskyrmionic nanodomains using  a phenomenological Ginzburg-Landau-Devonshire model based on the systematic symmetry-based free-energy  expansion and a simulated annealing method. 
In this way, an independent verification of the previous predictions based on the molecular dynamics with the atomistic model was obtained.
Moreover, the fact that the standard version on the discretized Ginzburg-Landau-Devonshire model is general enough to include the antiskyrmionic solutions also sheds a new light on the proposed stabilization mechanism\cite{Goncalves2024}.
Last but not least, the availability of antiskyrmions within the phase-field approach gives an access to  timescales and lengthscales larger than those accessible in the calculations with the atomistic models.
In this work, we have exploited this advantage of the phase-field approach and explored in more detail the electric field driven process of formation and annihilation of antiskyrmions as well as the targeted transport of antiskyrmionic ferroelectric bubble domains.

\section{Model and methods}

The phase-field approach consists in a numerical simulation of the energy-optimizing relaxation process for the spatially varying field of the electric polarization, subject to the dissipative Landau-Khalatnikov dynamics and  the direct interaction with the homogeneous and inhomogeneous strain fields, described by various interaction terms in a given Ginzburg-Landau-Devonshire  functional. 
The analytical definition  of the model employed in the present work was fully described in Refs\,\cite{Hlinka2006, Marton2010}. 
For the sake of comparison with our previous work, we have adopted Landau-Devonshire model parameters from Refs.\,\onlinecite{Stepkova2012,Stepkova2015}, and so it is the model B of Refs.\,\onlinecite{HlinkaBook2016} with the temperature fixed to 118\,K and $q_{1212}=1.57$\,nJm/C$^{2}$.
The only difference with Refs.\,\onlinecite{Stepkova2012,Stepkova2015,HlinkaBook2016} is the 4 times smaller magnitude of the gradient coefficients used in the present work, so that $G_{1111} = 127.5$\,pJm$^{3}$/C$^{2}$, $G_{1122} = -5$\,pJm$^{3}$/C$^{2}$, and $G_{1212} = 5$\,pJm$^{3}$/C$^{2}$.
Reduction of the gradient terms was primarily motivated by the enhancement of the lattice pinning and the earlier comparisons of ab-initio and phase-field domain wall profiles\cite{Taherinejad2012}, which had already indicated that the gradient coefficients estimated  from the experiments in the cubic phase of barium titanate are not appropriate for a quantitative description of thicknesses of domain walls in its low-temperature rhombohedral phase.

The simulations were carried out with a discrete time step of 1.25\,ps under a constant volume and shape of the supercell,  the latter corresponding to the mechanically free monodomain state with the spontaneous polarization along the pseudocubic [111] direction. 
The [111]-polarized monodomain state was used as the initial configuration.
We have typically worked with a rectangular supercell with dimensions {\(32\times32\times32\) nm\(^3\)} along the  crystallographic axes of the parent cubic perovskite phase. 
The polarization field was discretized on a regular mesh with a spatial step of 0.5\,nm and subject to  periodic boundary conditions.
The results are presented in a symmetry-adapted orthonormal coordinate system $\{rst\}$ carried by the orthogonal set of unit vectors  ${\bf r}\parallel [111]$,  ${\bf s}\parallel [\bar{1}10]$, and  ${\bf t}\parallel [\bar{1}\bar{1}2]]$, similarly as in Refs.\,\onlinecite{Marton2010,Stepkova2012, Stepkova2015}.

\section{Creation of antiskyrmion by inhomogeneous fields}

In contrast to the former work\cite{Goncalves2024}, where antiskyrmions were obtained by relaxing from an existing initial inverted antiparallel ferroelectric nanodomain of a hexagonal or triangular cross-section, 
 in this work we have created the antiskyrmions by various local electric fields, comparable to the homogeneous thermodynamical coercive field value  for 180-degree switching, which is about 
 60\,MV/m for the adopted Landau-Devonshire model.
 
\begin{figure}
\centering
\includegraphics[width=.9\columnwidth]{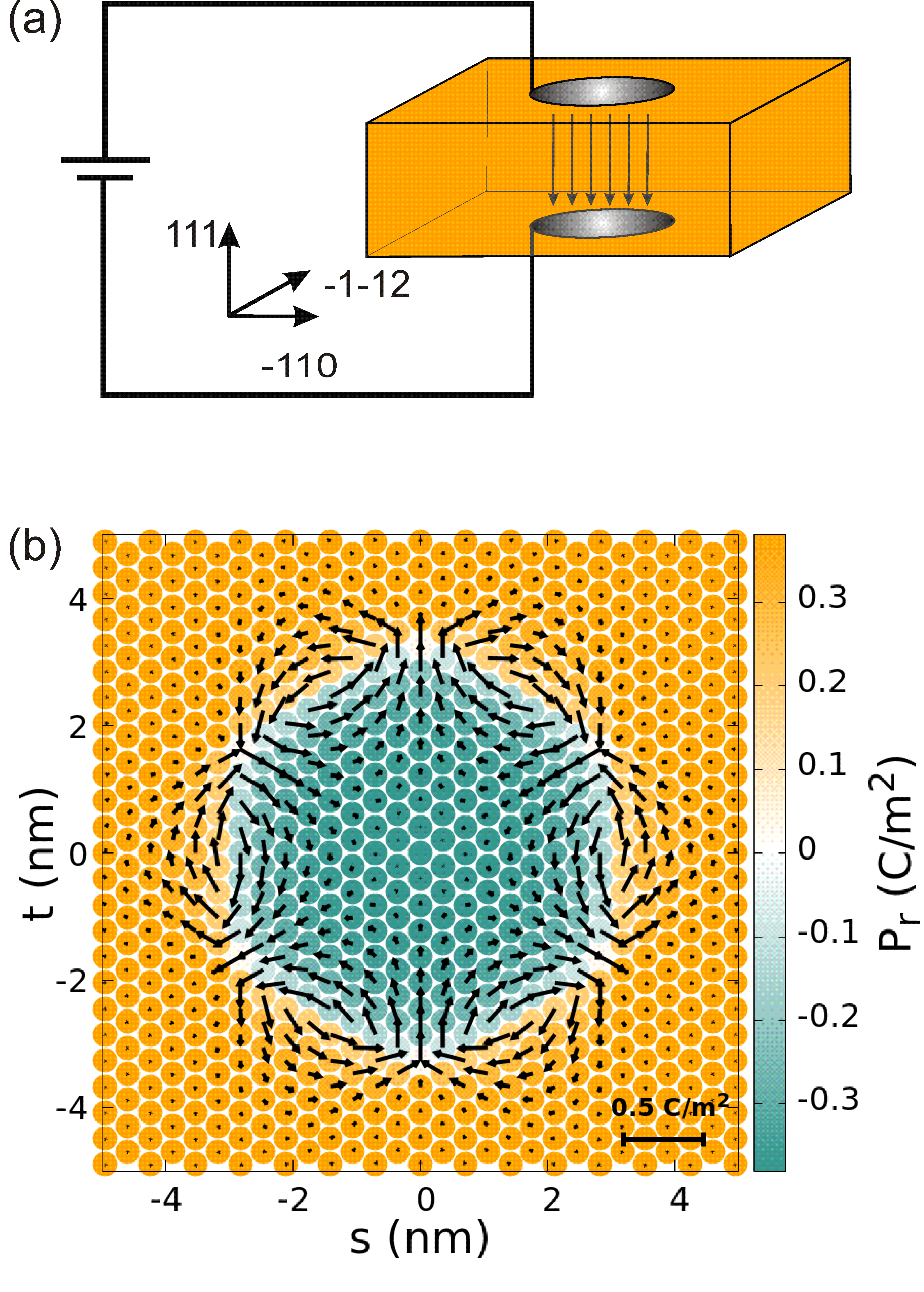}
\caption{ Creation of the antiskyrmion by electric field. (a) Schema showing the pseudocubic crystallographic axes and the orientation of the locally homogeneous field used to create the antiskyrmion. (b) Cross-section through the stable antiskyrmion nanodomain formed after the application and removal of the 80\,MV/m field, homogeneous within an infinite tube of 6\,nm diameter. The arrows stand for the in-plane polarization, the segment in the corner indicates the in-plane magnitude scale, colors show the out-of-plane component.
}
\label{fig:obr1}
\end{figure}

 For example, we applied a writing field, of 80\,MV/m, oriented opposite to the background monodomain polarization, within an infinitely long cylinder of a 6\,nm diameter, as schematically shown in Fig.\,\ref{fig:obr1}a.
This writing field reversed
the polarization inside of the cylinder by 180 degrees within a fraction of ps. 
In the next few ps, the non-collinear texture was spontaneously formed around the curved  boundary of this nanodomain.
This texture had the same qualitative geometrical and topological characteristics as  the antiskyrmions reported in the shell-model molecular dynamics study of the Ref.\,\onlinecite{Goncalves2024}.
The nanodomain and its antiskyrmionic character persists also when the field was removed. 
The final pattern with a core diameter of about 7\,nm is shown in Fig.\,\ref{fig:obr1}b. 
This diameter corresponds to 17 spatial steps of the discrete grid 
in the $s$-direction, so that in terms of definitions adopted in Ref.\,\onlinecite{Goncalves2024}, it is $N=17$ antiskyrmion. 

The same regular antiskyrmions could be created by axially symmetric field with a Gaussian spatial profile.
For example, the field with a full width at half maximum of 4\,nm and the peak value of 175\,MV/m applied during 0.1\,ps  also created the
7\,nm-diameter antiskyrmion shown in Fig.\,\ref{fig:obr1}b.
When the field was applied longer, the final diameter increased.
For example, 1\,ps exposure to the same Gaussian field resulted in the $N=23$ antiskyrmion, with a diameter of about 9\,nm.
With the adopted model, the smallest stable antiskyrmion at zero field was that with $N=13$, while in the previous atomistic shell model calculations,
conducted at $T=1$\,K, the smallest antiskyrmion was that with $N=9$.
Therefore, we have carried out also additional calculations with the Landau potential corresponding to $T=1$\,K, and verified that there the smallest stable antiskyrmion has $N=11$, a closer value to the previous atomistic theory prediction\cite{Goncalves2024}.
%

\section{Annihilation of the antiskyrmions}

The opposite process, that of the antiskyrmion contraction and its irreversible collapse, can be  performed  by an electric field applied parallel to the background polarization.
The orientation of this field  penalizes the core polarization and favors that of the background.
Therefore, this field promotes a contraction of the antiskyrmion radius while keeping its macroscopic $3m$ symmetry.
Since the skyrmionic states are locally stable, the  field needed to start this process needs to overcome a threshold value.
This value is less than one order of magnitude lower than the thermodynamic coercive field\cite{Goncalves2024}.
Therefore, the erasing field can be homogeneous throughout the entire simulation box.
We have numerically determined that the threshold homogeneous field needed to initiate the process of the irreversible contraction  of the $N=17$ antiskyrmion in the present model is about 2.4\,MV/m.

The antiskyrmion annihilation driven by a 10\,MV/m homogeneous field is documented in Fig.\,2.
Although the simulated temporal evolution is a continuous process, it proceeds by a succession of faster and slower episodes, the latter implying passages  through several distinct quasistationary states.
These quasistationary states, enumerated in Fig.\,2 by sequential numbers 0-4, correspond to symmetric antiskyrmionic configurations with odd size numbers $N=17,15,13,11$ and 9, respectively.
Let us note that the [111]-projection of the three-dimensional  discrete grid forms a two-dimensional lattice of points in which the local polarization is evaluated, and that the nearest neighbors form elementary equilateral triangular plaquettes (see Fig.\,2.).
We can see that the odd values of $N$ permit simultaneous location of the axis of the antiskyrmion on the lattice point and the six circumferential vortexes at the centers of the elementary triangular plaquettes.
This clearly indicate that pinning of vortex cores to the discrete lattice is essential for the stability of  antiskyrmions. 

In addition to the in-plane polarization components, the color circles depicts the invariant skyrmion density\cite{Goncalves2024}.
We can see that  the vortex centers in the stationary states coincide with the sharp hot spots of the skyrmion density.
As expected, in all (b)-(f) panels, the skyrmion density integrates to the overall topological charge of minus two.
During the passage between subsequent stationary states, the overall topological charge was preserved.
The integral topological charge vanished abruptly only when the diameter was so small that the polarization at the center of the core itself vanished.
This instant is indicated  in Fig.\,2a by a vertical dashed line.

\begin{figure}
\centering
\includegraphics[width=.9\columnwidth]{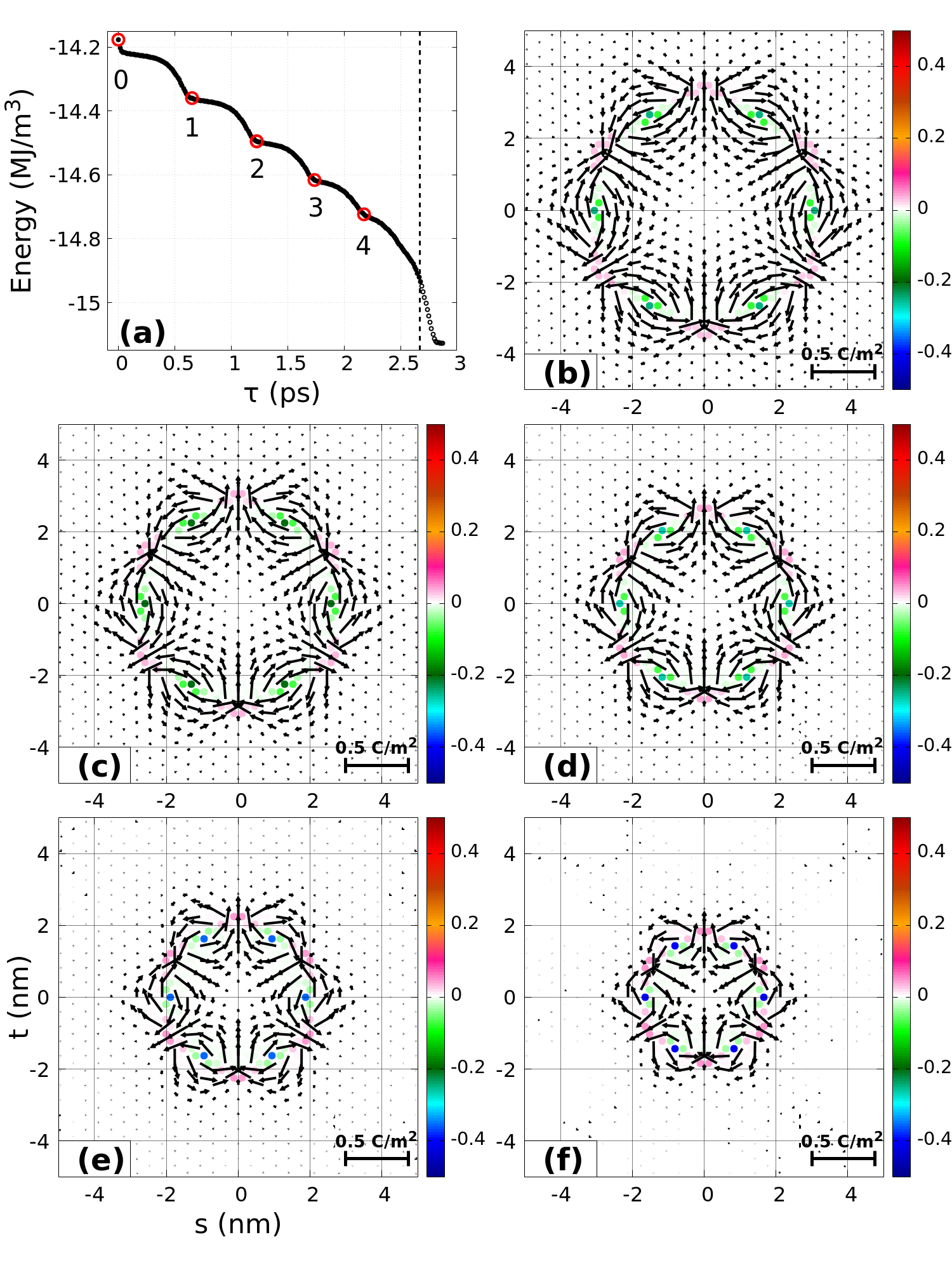}
\caption{ 
Annihilation of the $N=17$ antiskyrmion by a 10\,MV/m homogeneous field, parallel to the background polarization.
(a) Time evolution of the overall average elastic Gibbs free energy density during the field-induced process of the antiskyrmion collapse.
Numbers 0-4 within the panel indicate the temporal location  quasi-stationary configurations with $N=$\,17, 15, 13, 11 and 9.
The vertical line indicates the instant of the topological charge collapse ($-2 \rightarrow 0$).
(b)-(f) Energetically favored transient configurations corresponding to minima 0-3 and the inflection point 4 on panel (a) are  projected on the $st$ plane.
Arrows show the in-plane components of polarization, while colors stand for the invariant skyrmion density values calculated at each elementary triangular plaquette within the [111]-projection of the three-dimensional discrete simulation grid. 
}
\label{fig:obr2}
\end{figure}

\section{Transport of antiskyrmions by electric field}
Finally, we have explored the possibility to move the antiskyrmion in a desired direction by suitably tailored electric fields. 
We have observed  that a simple spatial translation of the writing field described above can lead to the formation  of various elongated nanodomains with irregular or significantly deformed shape.
One of the efficient method allowing  to  transport  the antiskyrmion without much shape distortion  consisted in the usage of the bipolar electric field, depicted in Fig.\,3.

This  field was a superposition of two electric field distributions, a stronger ($E_- \approx 52$\,MV/m) writing field, oriented opposite to the background polarization,  and a weaker ($E_+ \approx 18$\,MV/m) erasing field, pointing in the same direction as the background polarization.
The writing and the erasing fields were both homogeneous within the circular diameter 
adapted to the antiskyrmion of interest (in present case, 6\,nm), but they were mutually shifted   
by about 1.5-2\,nm.
Resulting electric "tweezer" field enabled to shift the antiskyrmion core center towards the center of the writing field.
When such tweezer field is translated in space with a velocity slower than about 500\,m/s, the antiskyrmion can travel together with the field over an arbitrary distance (see fig\,4).
By changing the direction of the mutual shift between the two fields, the antiskyrmion could be equally well dragged along any other direction in the $st$ plane.

\begin{figure}
\centering
\includegraphics[width=.9\columnwidth]{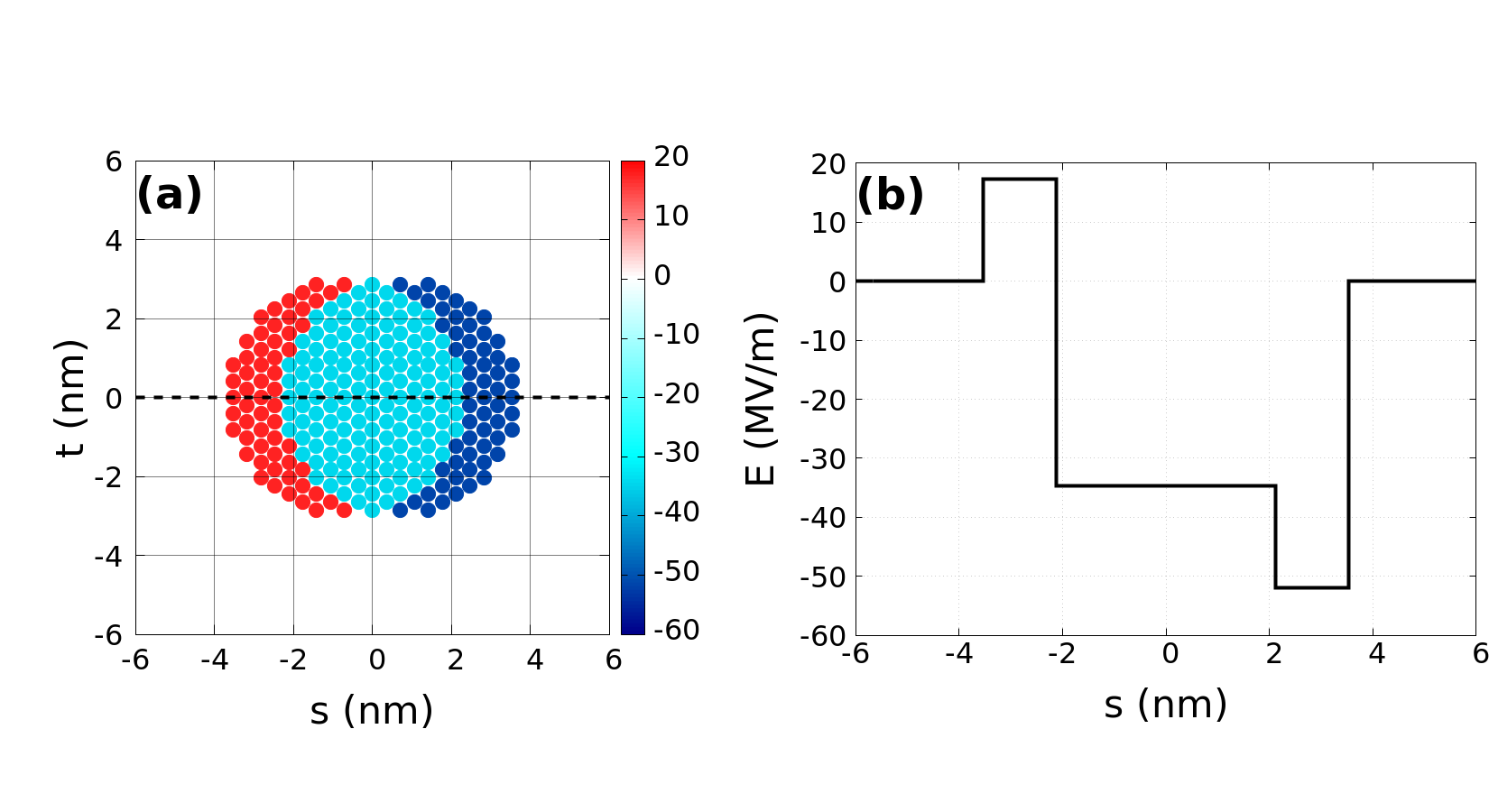}
\caption{Suitable electric tweezer field enabling the antiskyrmion transport in the  positive sense of $s$-axis (to the right). (a) Planar cross-section of the field in the $st$ plane and (b) its profile  along the  dashed line indicated on  (a).
}
\label{fig:obr4a}
\end{figure}

\begin{figure}
\centering
\includegraphics[width=.9\columnwidth]{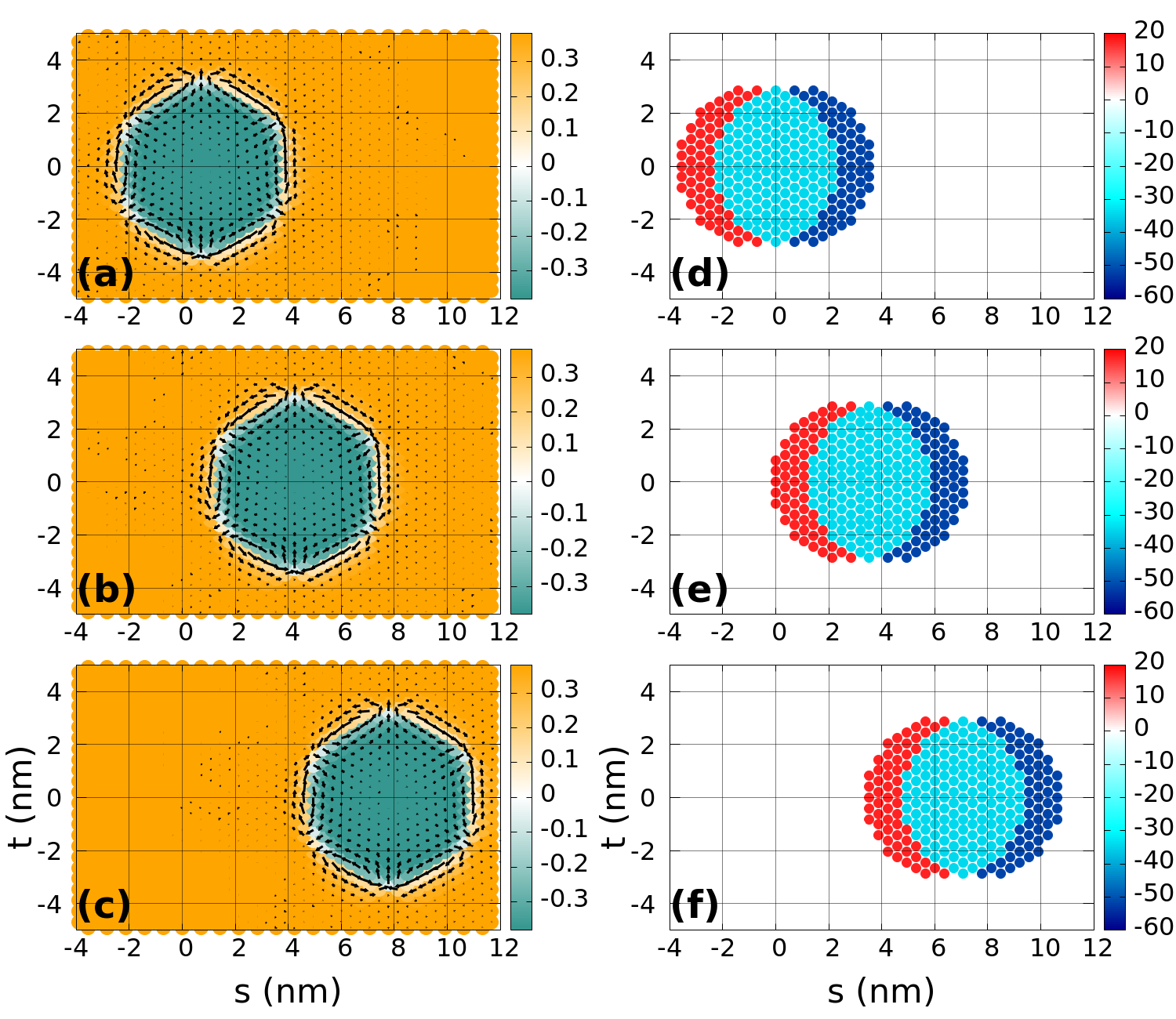}
\caption{
Translation of the antiskyrmion by "tweezer" electric field of Fig.\,3 in the direction $s$. 
Panels (a), (b) and (c) show polarization at the at three subsequent instants of the simulation, while the corresponding drifting electric tweezer field is shown in patels (d), (e) and (f). 
The color code in panels (a), (b) and (c) indicates $P_{\rm r}$ component of the polarization in C/m$^2$ and in panels (d), (e) and (f) it gives the $E_{\rm r}$ field component in MV/m.
}
\label{fig:obr4b}
\end{figure}

\section{Discussion and conclusion}

In summary, the present phase-field calculations confirmed the existence of the string-like antiskyrmionic bubble domain solutions in the discretized Ginzburg-Landau-Devonshire model.
The adopted Landau-Devonshire potential had standard parameters for rhombohedral barium titanate. 
The resulting antiskyrmion patterns were qualitatively similar to those encountered in the original atomistic simulations\cite{Goncalves2024}.
In particular, the polarization pattern is composed from sextuplet of polar vortexes, it has an overall $3m$ symmetry, and the overall invariant skyrmion topological charge is minus two.
From this perspective, these low-temperature ferroelectric bubble domains are very different from lead-titanate ones\cite{Gomez-Ortiz2024,Aramberri2024}. 

Present simulations focused on several field-induced processes.
The spontaneous phase of the antiskyrmion formation process was activated by application of a focused electric field impulse.
Typically, the above-coercive field creates a narrow beam of the inverted polarization, which is subsequently decorated by the characteristic topologically nontrivial, noncollinear polarization pattern. 
The transport of antiskyrmionic nanodomains was achieved by a suitably designed bipolar tweezer field that combines off-centered writing and erasing fields of opposite sense.
The analysis of the process of the irreversible annihilation of the antiskyrmion by a homogeneous sub-coercive field revealed a sequence of stationary states possessing odd values of the dimensionless diameter $N$.
In overall, the reported annihilation process provides an instructive example of a topological collapse in an already energetically destabilized transient state. 

The above findings confirms that the discrete lattice pinning of vortex centers plays the essential role in the stability of these antiskyrmionic nanodomains.
Therefore, it can be anticipated that the presence of thermal fluctuations will further facilitate the transport and annihilation of the antiskyrmions at finite temperatures.
At the same time, some of the numerical values obtained in this study, such as the critical anstiskyrmion diameters or the threshold field magnitudes, are expected to be considerably sensible to the model parameters and in particular, to the magnitude of the gradient tensor components and to the size of the discrete spatial step.
Similarly, the temporal rates should be considered with a caution since the value of the kinetic parameter was derived from a very simple assumption and from the room-temperature relaxation time\cite{Hlinka2007}. 

Nevertheless, we did not aimed here to approach real barium titanate properties by a tight fine-tuning of all these parameters.
We believe that in the near future, a more precise quantitative  predictions for the nanoscale phenomena in the low-temperature of barium titanate will become available in a more straightforward and convincing manner  from the recently developed ab-initio based effective Hamiltonians\cite{Mayer2022, Nishimatsu2008, Marathe2017}, from the systematic second-principles models\cite{Zhang2023, ScaleUp}, or also from other carefully designed and tested interatomic potentials\cite{QiY2016}.
Still, 
the present phase-field calculations should give a realistic qualitative picture of antiskyrmion formation, sliding and annihilation processes.
Most likely the present results are also indicating experimentally relevant orders of magnitude of timescales and lengthscales of these processes.
We believe that these findings will stimulate further research of this  type of interesting topological ferroelectric bubble domains.
%

This work was supported by the Czech Science Foundation (project no. 19-28594X). Authors acknowledge multiple inspiring scientific exchanges with Mauro A.P. Gonçalves, Pavel Márton  and Marek Paściak from FZU and with Florian Mayer from MCL Leoben.

\end{document}